\documentclass[conference]{IEEEtran}
\usepackage{graphicx}
\usepackage{amsmath}
\usepackage{booktabs}
\usepackage{xcolor}

\ifCLASSINFOpdf
\else
\fi
\usepackage{float}
\usepackage{setspace}
\usepackage[top=0.8in, bottom=1.1in, left=0.63in, right=0.63in]{geometry}
\begin{document}
\title{Comparison of Optimizers for Fault Isolation and Diagnostics of Control Rod Drives}

\author{\IEEEauthorblockN{Ark Ifeanyi}
\IEEEauthorblockA{University of Tennessee\\ Knoxville, Tennessee 37996, USA\\
Email: aifeanyi@vols.utk.edu}
\and
\IEEEauthorblockN{Jamie Coble}
\IEEEauthorblockA{University of Tennessee\\ Knoxville, Tennessee 37996, USA\\
Email: jamie@utk.edu}
% \and
% \IEEEauthorblockN{James Kirk\\ and Montgomery Scott}
% \IEEEauthorblockA{Starfleet Academy\\
% San Francisco, California 96678-2391\\
% Telephone: (800) 555--1212\\
% Fax: (888) 555--1212}
}


%Telephone: (800) 555--1212, Fax: (888) 555--1212}


\maketitle

\begin{abstract}
%\boldmath
This paper explores the optimization of fault detection and diagnostics (FDD) in the Control Rod Drive System (CRDS) of GE-Hitachi's BWRX-300 small modular reactor (SMR), focusing on the electrically powered fine motion control rod drive (FMCRD) servomotors. Leveraging the coordinated motion of multiple FMCRDs for control rod adjustments, the study proposes a deep learning approach, utilizing one-dimensional convolutional neural network (1D CNN)-based autoencoders for anomaly detection and encoder-decoder structured 1D CNN classifiers for fault classification. Simulink models simulate normal and fault operations, monitoring electric current and electromagnetic torque. The training of the fault isolation and fault classification models is optimized. Various optimizers, including Adaptive Moment Estimation (Adam), Nesterov Adam (Nadam), Stochastic Gradient Descent (SGD), and Root Mean Square Propagation (RMSProp), are evaluated, with Nadam demonstrating a relatively superior performance across the isolation and classification tasks due to its adaptive gradient and Nesterov components. The research underscores the importance of considering the number of runs (each run has a different set of initial model parameters) as a hyperparameter during empirical optimizer comparisons and contributes insights crucial for enhancing FDD in SMR control systems and for the application of 1D CNN to FDD.
\end{abstract}

% \fi
\IEEEpeerreviewmaketitle

\section{Introduction}
\label{sec:intro}

Nuclear reactors play a pivotal role in meeting the world's energy demands, and the efficient regulation of thermal power in reactors is critical for their safe and optimal operation. Control rods (CRs) are fundamental components in nuclear reactors, modulating the rate of fission by absorbing neutrons when inserted into the core. This fission rate in turn controls thermal power output. The control of the CR position (insertion/withdrawal) is facilitated by control rod drives (CRDs), a mechanism vital for maintaining the stability and performance of the reactor core.

\begin{figure}[t]
\centering
\includegraphics[width=0.47\textwidth]{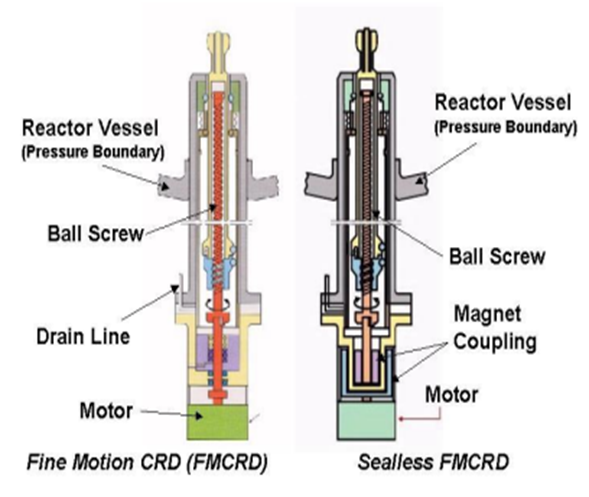}
\caption{Fine Motion Control Rod Drive Systems}
\label{fmcrd_systems}
\end{figure}

In the evolution of reactor designs, the paradigm is shifting towards Small Modular Reactors (SMRs) due to their compact size and operational flexibility. Notably, GE-Hitachi's BWRX-300 SMR design represents a significant advancement by replacing conventional hydraulically-driven CRDs with electrically powered fine motion control rod drives (FMCRDs). This FMCRD (see Fig. \ref{fmcrd_systems}) enhances control precision and responsiveness, contributing to the adaptability and efficiency of the reactor.

Within the reactor core, CRs operate in banks, collectively influencing the power profile. Currently, one challenge in monitoring FMCRD conditions is that the movement of CRs happens intermittently. This means that most of the time, CRs remain still, only moving when commanded due to changes in reactor conditions. Additionally, when CRs are stationary, the servomotor, which controls their movement, doesn't generate any data that could help assess their condition. On the other hand, the coordinated motion of multiple FMCRDs powered by servomotors within these banks creates a unique opportunity for monitoring and assessing the health of individual components. Leveraging this coordinated operation, an approach to fault detection and diagnostics (FDD) that capitalizes on the synchronized behavior of servomotors within a bank was proposed in a previous work \cite{ifeanyi2024deep}. The paper monitored the torque and position of the FMCRD to detect and diagnose three fault types: short-circuit fault, ball-screw jam fault, and ball-screw wear fault. 

This current research aims to extend the previous research by achieving two primary objectives: first, to improve detection and diagnostics results by monitoring an additional property of the system (current); second, to systematically investigate the effect of four different first-order optimizers on these FDD tasks. By addressing these objectives, the study not only contributes to the understanding of FDD methodologies in the context of FMCRD servomotors but also provides valuable insights into the optimization of deep learning models for enhanced performance in fault diagnosis within nuclear reactor systems. 

While SMRs are garnering considerable attention and support for their promising features, operationalizing these reactors in the United States is still on the horizon, with the first deployment anticipated in 2029 \cite{us_smr_certification_apnews, nrc_certifies_smr_design}. The FMCRDs, being integral to the operation of GE-Hitachi's BWRX-300 SMR, serve as a backbone for the reactor's flexibility and efficiency. The ease and accuracy with which faults in these motors can be detected are critical factors influencing the economic feasibility of these SMRs. Effective fault detection not only ensures the safety and reliability of SMRs but also has the potential to significantly reduce operation and maintenance (O\&M) costs, thereby enhancing the contribution of the global nuclear fleet to energy security.

Our approach involves the application of deep learning to the FDD tasks. Investigating the best Keras optimizer to employ in its default state when applying deep learning to fault detection and diagnostics is a crucial step in the model development process. Default optimizers in Keras, such as Adam, RMSprop, or SGD, come with pre-defined hyperparameters that are generally well-suited for a wide range of tasks. This default state exploration is particularly valuable in scenarios where computational resources are limited, as optimizing the hyperparameters of the optimizer can be computationally expensive. Moreover, the process of tuning optimizer hyperparameters demands a profound understanding of the specific problem domain, and the optimal configuration may vary across different fault detection tasks. By initially assessing the default optimizers, practitioners can gain insights into their model's performance without the need for extensive hyperparameter tuning, allowing for a more efficient and focused optimization process. This pragmatic approach is especially relevant in fault detection and diagnostics applications, where the emphasis lies on promptly and accurately identifying deviations from normal system behavior, and a well-chosen default optimizer can often provide a strong starting point for model development.

The subsequent section (section \ref{sec:lit_review}) of this paper delves into a thorough examination of existing research on fault detection in similar systems and establishes the foundational knowledge for our investigation. Following this, section \ref{sec:methods} meticulously details our approach, utilizing Simulink models and deep learning techniques to simulate and analyze servomotor operations. Section \ref{sec:results} presents findings from the experiments, offering insights into the effectiveness of various optimizers in fault detection and diagnosis. Finally, in section \ref{sec:conclusion}, the paper concludes with a thoughtful synthesis of the outcomes and their implications in the realm of Small Modular Reactors, emphasizing the significance of our contributions to the broader discourse on nuclear energy system reliability and operational efficiency.  

\section{Literature Review}
\label{sec:lit_review}
FDD encompasses the detection of faults, mapping the detected faults to their origins (isolation), and identifying the type of fault detected (diagnostics) \cite{fdd2011}. In this research, the focus is on isolation and diagnostics. Although some researchers have explored FDD within control rod drives (CRDs), which are primarily used for safety actuation in the case of reactor SCRAM, \cite{crds_review}, there's a notable gap in understanding FMCRDs, used for normal insertion and withdrawal to manage power profiles across the reactor core. One prior study on fault detection in FMCRDs employed displacement monitoring and a statistical thresholding approach. This involved setting a threshold on the average difference between demanded and actual rod positions within a bank \cite{withinBankComp}. The expected variation in this average difference was meant to distinguish between healthy and faulty banks \cite{withinBankComp}. However, this specific attempt only successfully detected one fault type, namely, short-circuit. Another recent study applied principal component analysis (PCA) to anomaly detection in FMCRDs, using T-squared statistics and Q-statistics for this purpose \cite{aifeanyi_within_bank_cond}. This second study extended the observation period to include more data points in the training data to enable detection and isolation. In contrast, the approach presented in this paper aims to capture intricate relationships between variables, potentially enabling the detection of various fault types with a reduced number of observations in the training data and a limited number of samples. Additionally, this paper contributes to the FDD cycle by addressing fault isolation, and diagnostics for this significantly underrepresented system in the existing literature. 

Various anomaly detection techniques have been employed over the years, encompassing classification-based, clustering-based, nearest neighbor-based, and statistical approaches \cite{anomalyDetect}. In the context of this paper, fault detection involves classifying a bank of control rods as either faulty or healthy. Anomaly detection can be achieved through both traditional machine learning techniques and more advanced deep learning methods \cite{deepLearnDetect}. In the realm of deep learning, autoencoders stand out as widely utilized architectures for anomaly detection and fault isolation, particularly in the domain of prognostics and health management (PHM) \cite{rezaeianjouybari2020deep}. Autoencoder reconstruction has proven effective in detecting anomalies in industrial motors \cite{autoencoderRecon}, while variational autoencoders have been successfully applied to identify anomalies in electric drives \cite{unsupervisedAnomaly}. Given their widespread acceptance and success in similar applications \cite{autoencoder1, autoencoder2}, a variant of the autoencoder was selected as the architectural choice for fault detection and isolation in this work.

Concerning the diagnostic assessment of motors, various approaches have been historically employed, with motor current signature analysis (MCSA) standing out as the most prevalent \cite{mcsa2007motor,mcsa2}. MCSA involves scrutinizing the properties of the motor's current signal and aligning them with established behaviors for different fault types to classify the investigated fault \cite{mcsa3}. However, a notable drawback of this method is its dependency on previously established behaviors, making fault classification challenging in the absence of such references. Moreover, similar characteristics among different faults increase the risk of misdiagnosis. Additionally, the approach relies on monitoring motor current, rendering classification unfeasible in cases where current signals are not observed. Addressing these challenges, recent techniques have turned to the application of neural networks in induction motor fault diagnostics \cite{classif2003neural}. In these methods, signals like vibrations may be explored for diagnostic purposes. Some researchers have also explored motor fault diagnosis using Fuzzy Logic-based techniques, either as precursors to other diagnostic methods or as the primary diagnostic algorithm \cite{diag2000review}. However, limited work has been found for diagnostics in FMCRDs, and to the best of our knowledge, no known published work has discussed their application in within-bank analysis. It is not uncommon to find anomaly detection in tandem with diagnosis in various fields including in the health sector where diagnosing heart disease has been treated as a hierarchical problem \cite{wang2023hierarchical}. First, an autoencoder is used to detect anomalies in electrocardiogram (ECG) signals before transferring the knowledge from the trained autoencoder to a classifier for diagnosis \cite{wang2023hierarchical}. In this current study, no knowledge transfer was done and a 1D-CNN-based classification model was employed to attempt diagnostics within banks of FMCRDs. 1-dimensional convolutional neural network (1D-CNN)-based autoencoders were used in this study instead of typical fully connected layer-based autoencoders to maintain the temporal correlations in the input data. To add to that, utilization of 1D-CNNs has proven effective for fault diagnosis in scenarios where the input signals are sequences \cite{1d_cnn_1,1d_cnn_2,1d_cnn_3}, justifying their application throughout this paper.

A specially designed deep learning classification model was used for fault diagnostics. One benefit of deep learning models is that they may reveal complicated correlations without necessarily needing feature engineering \cite{ahmed2023deep}. The constrained coding layer also helps autoencoders extract valuable information, which is important for the application under investigation because of the servomotors' sporadic operations explained in section \ref{subsec:data} and the small number of samples available. One significant benefit of the suggested technique over the previously discussed statistical methods is its capacity to concurrently take into account the time dependency of each sensor variable and the nonlinear correlations between numerous sensor variables.

Moving away from FDD, optimization algorithms determine the speed of convergence and the predictive performance of models. They do this by updating the parameters of the model using different update rules. The update rule uses historical values of parameters and gradients of the loss function, as well as the hyperparameters of the optimization algorithm, to calculate the model’s parameters at the current timestep \cite{wilson2017marginal, schneider2019deepobs}. The four investigated optimizers and their update rules are shown in Table \ref{optimizer_update_rules}. Stochastic gradient descent (SGD) optimizes parameters using the negative gradient and a learning rate \cite{sgd}. Root mean square propagation (RMSProp) adapts learning rates per parameter with moving averages of squared gradients \cite{rmsprop} whereas Adam adds momentum, decaying averages of past gradients, and bias correction \cite{2014adam}. Finally, Nadam combines Nesterov’s momentum with Adam's adaptive learning rates for enhanced stability and convergence \cite{Nadam}.

It is worth mentioning that first-order optimizers when optimized, exhibit inclusion hierarchy relationships where some optimizers can approximate others as shown in equations \eqref{eq:subeq1}, \eqref{eq:subeq2}, and \eqref{eq:subeq3}. These relationships mean that the more specific optimizers like SGD, when optimized, should not outperform the more general optimizers like RMSProp, Adam, and Nadam in their optimized states \cite{choi2019empirical}. The challenge is that the general optimizers can be more expensive to optimize because of their increased number of hyperparameters so achieving this relationship might be impractical for most deep learning practitioners \cite{choi2019empirical}. As a result, the default Keras versions of all the optimizers were compared in this research for application in FDD of FMCRDs.

\begin{subequations}\label{inclusion_hierarchy}
    \begin{align}
        \text{SGD} &\subseteq \text{Momentum} \subseteq \text{RMSProp} \label{eq:subeq1}\\
        \text{SGD} &\subseteq \text{Momentum} \subseteq \text{Adam} \label{eq:subeq2}\\
        \text{SGD} &\subseteq \text{Nesterov} \subseteq \text{Nadam} \label{eq:subeq3}
    \end{align}
\end{subequations}

\begin{table}[t!]
    \centering
    \caption{Mathematical Update Rules for Keras Optimizers}
    \label{optimizer_update_rules}
    \begin{tabular}{l l}
        \toprule
        \textbf{Optimizer} & \textbf{Update Rule} \\
        \midrule
        SGD & $w_{t+1} = w_t - \eta \nabla J(w_t)$ \\
        Adam & $m_{t+1} = \beta_1 m_t + (1 - \beta_1) \nabla J(w_t)$ \\
        & $v_{t+1} = \beta_2 v_t + (1 - \beta_2)(\nabla J(w_t))^2$ \\
        & $w_{t+1} = w_t - \frac{\eta}{\sqrt{v_{t+1}} + \epsilon} m_{t+1}$ \\
        RMSProp & $v_{t+1} = \beta v_t + (1 - \beta)(\nabla J(w_t))^2$ \\
        & $w_{t+1} = w_t - \frac{\eta}{\sqrt{v_{t+1}} + \epsilon} \nabla J(w_t)$ \\
        Nadam & $m_{t+1} = \beta_1 m_t + (1 - \beta_1) \nabla J(w_t)$ \\
        & $v_{t+1} = \beta_2 v_t + (1 - \beta_2)(\nabla J(w_t))^2$ \\
        & $m_{\text{corr}} = \frac{\beta_1 m_{t+1} + (1 - \beta_1) \nabla J(w_t)}{1 - \beta_1^{t+1}}$ \\
        & $w_{t+1} = w_t - \frac{\eta}{\sqrt{v_{t+1}} + \epsilon} m_{\text{corr}}$ \\
        \bottomrule
    \end{tabular}
\end{table}

Table \ref{optimizer_update_rules} provides the mathematical update rules for popular first-order optimizers: Stochastic Gradient Descent (SGD), Adam, RMSProp, and Nadam. Here, $w_t$ represents the model weights, $\eta$ is the learning rate, $\nabla J(w_t)$ is the gradient of the loss function, and $\epsilon$ is a small constant for numerical stability. The parameters $\beta_1$ and $\beta_2$ control the exponential decay rates for the moving averages in Adam and Nadam.

\section{Methodology}
\label{sec:methods}
This research employs a Simulink model to simulate the operations of FMCRD servomotors under normal conditions and various fault modes. Monitoring the electric current in one of three available phases and the electromagnetic torque of these servomotors, we utilize deep learning models to isolate and diagnose faults, encompassing tasks such as anomaly detection and fault classification. The efficacy of different optimizers, including Adaptive Moment Estimation (Adam), Nestrov Adam (Nadam), Stochastic Gradient Descent (SGD), and Root Mean Square Propagation (RMSProp), will be systematically evaluated. Key metrics, such as minimum loss value and convergence rate, will guide the selection of the most suitable optimizers for each specific task, contributing to the robustness and accuracy of the fault detection system.

\subsection{Data}
\label{subsec:data}
\begin{figure}[t]
\centering
\includegraphics[width=0.45\textwidth]{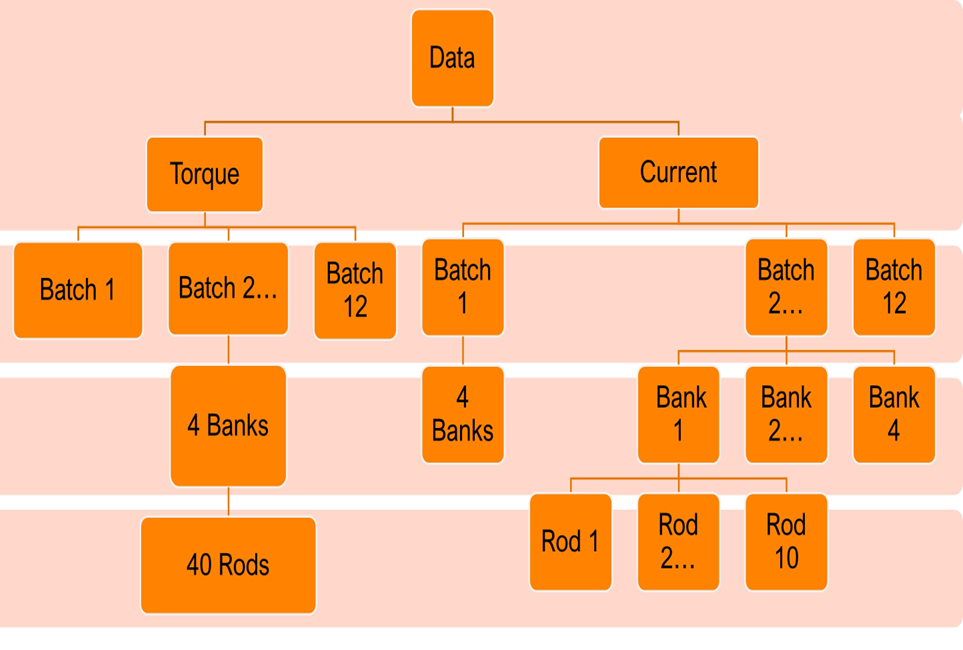}
\caption{Data Structure}
\label{data_structure}
\end{figure}

\begin{figure}[t]
\centering
\includegraphics[width=0.45\textwidth]{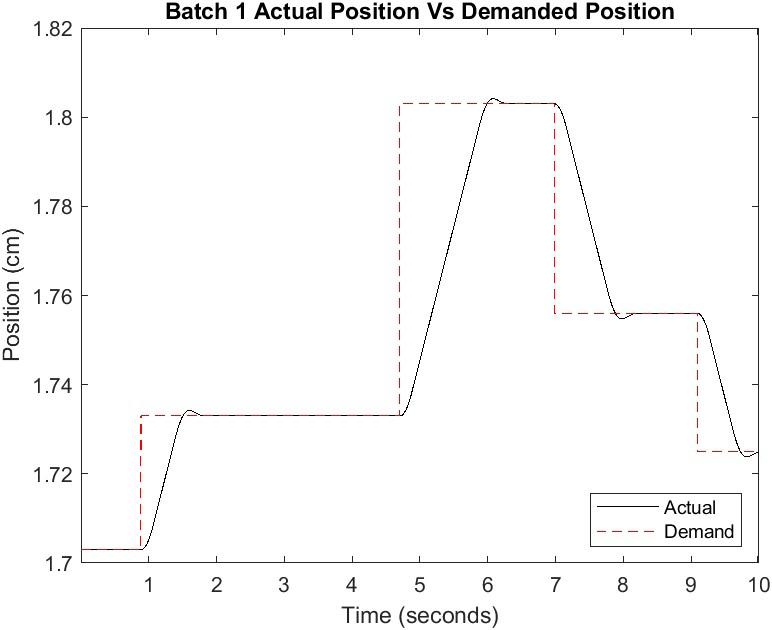}
\caption{Healthy position profile of a single batch}
\label{operations1}
\end{figure}

The data used in this study came from an FMCRD simulation model implemented in MATLAB/Simulink \cite{subramanian2023servomotor}. These data show how control rod banks react to demands for position changes during the course of a ten-second maneuver. Two properties of the FMCRD were monitored in this work; the first variable records the electromagnetic torques (in Newton-meters) of the bank's servomotors, and the second one tracks the electric current (in amperes) (see Fig. \ref{data_structure}). Twelve batches of data were generated for each variable, each displaying a unique motion profile over the observation time (see Fig. \ref{operations1}). Each batch has four banks: one bank that is in good health and three banks that are faulty, each of which represents a distinct tested fault type. Each bank comprises ten rods, with healthy banks containing all ten rods in good condition, and faulty banks having one faulty rod intentionally placed in position '10'. The data breakdown overview deliberately omits some components for summary purposes. The 'Torque' branch displays the overall count of rods and banks in any given batch, although it was specifically demonstrated for batch '2'. These sums are evenly distributed, ensuring that each bank, as exemplified in bank '1' of batch '2' in the 'Current' branch, consists of ten rods. The three investigated faults were short circuit, ball screw wear, and ball screw jam faults. The ball screw is a mechanical component of the drive system so the wear and jam faults are mechanical faults whereas the short-circuit fault is electrical. The wear fault was simulated as a gradual increase in mechanical load but the jam fault as an instant increment \cite{subramanian2023servomotor}.

Additionally, each rod produces five hundred thousand temporally related observations (excluding the starting point) for each measured property due to a sampling rate of 50kHz for the ten-second maneuvers. In this work, each bank, consisting of ten rods with ten-second sequences (500001 data points) of current or torque, is treated as one sample. The input shape is $(x,500001,10)$, where $x$ represents the number of banks. A total of forty-eight generated data banks were created, with twelve for each of the four categories: healthy (no-fault), short-circuit fault, ramp fault, and step fault.

For each task, the training and test sets have a distinct composition, which is explained in detail in the corresponding sections. For clarity, the defective rod was always located at the same variable index ('10') of the bank in all test scenarios, while it could theoretically be at any index and still cause the detection and isolation procedures to operate as intended. 

For brevity, this section specifically discusses the behavior of batch '1', although similar patterns were observed in the healthy banks of all other batches (see Fig. \ref{operations1}). The rods in that bank began to progressively rise one second after batch '1' was being monitored. This was the result of a request for a position modification. The rods overshot at the required point before stabilizing at the predetermined location, which was around 1.735 cm. Then, at the $5^{th}$ second, there was another ascent request, which resulted in another slow upward movement and a subsequent equilibrium settlement. The drive mechanism's operating periods are defined by these intervals during which the rods travel to reach the intended position. The fact that the servomotor only runs for a portion of the observation period means that there are few data points from the rod position signals and other monitored response signals that can be used to identify faults, making the FDD tasks more difficult. The step profile shows that the position demand is immediate, but the actual rod movements take time to unfold, therefore there is a temporal lag between the requested and actual locations.

\subsection{Fault Isolation}
\label{subsec:method_isol}
\begin{figure}[t]
\centering
\includegraphics[width=0.45\textwidth]{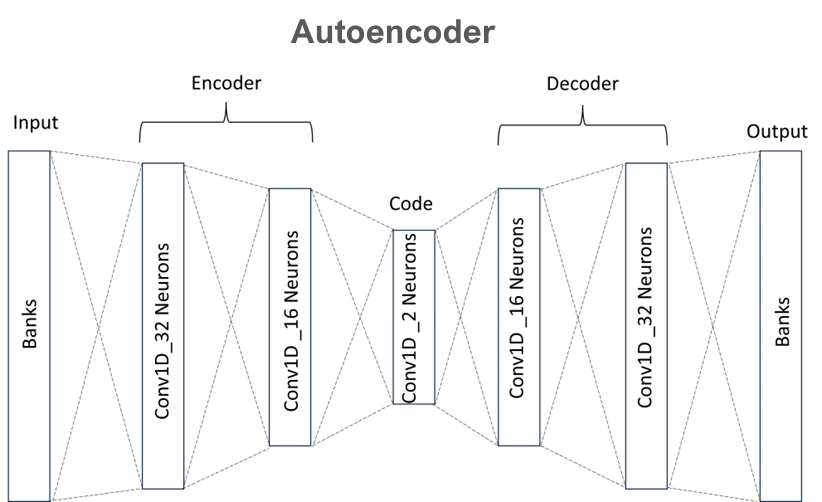}
\caption{1D-CNN autoencoder}
\label{autoencoder}
\end{figure}

In addressing the intricate challenge of fault isolation within banks of control rod drive servomotors, our research undertook a methodical exploration, leveraging advanced deep learning techniques. The initial pivotal step involved the selection of a first-order Keras optimizer, a critical influence on the subsequent training of our 1D-CNN autoencoder models (Fig. \ref{autoencoder}). These models, expected to be adept at capturing nuanced temporal dependencies within the electric current and electromagnetic torque signals of fine motion control rod drives (FMCRDs), formed the core of our fault isolation methodology.

The training phase commenced with datasets exclusively derived from healthy control rod drive banks (nine samples), enabling the models to learn and encode the subtle features inherent to normal operational conditions. This foundational training set the stage for subsequent fault isolation tasks, where deviations from the learned normalcy would serve as indicators of potential faults. To bolster the models' robustness, a separate set (three samples) of control rod drive banks in a healthy state was utilized for validation. This step, crucial for assessing generalization capabilities, aimed to ensure the methodology's effectiveness across diverse datasets representative of normal operational conditions.

Validation laid the groundwork for the establishment of a reconstruction error threshold, a pivotal parameter guiding the identification of anomalies during subsequent testing. Rigorous testing involved exposing the trained and validated models to datasets from control rod drive banks exhibiting various faults. The primary focus was on pinpointing rods with significantly higher reconstruction errors, indicative of potential deviations from the learned healthy operational patterns.

Recognizing the diversity of operational parameters and the potential impact of optimizers, our methodology embraced an iterative approach. The fault isolation task was systematically repeated for each first-order Keras optimizer, allowing for a comprehensive assessment of their influence on the performance of the 1D-CNN Autoencoder models. Furthermore, this process was replicated for both electric current and electromagnetic torque properties, providing a nuanced understanding of fault isolation across distinct operational parameters.

\subsection{Fault Diagnostics}
\label{subsec:method_diag}
\begin{figure}[t]
\centering
\includegraphics[width=0.45\textwidth]{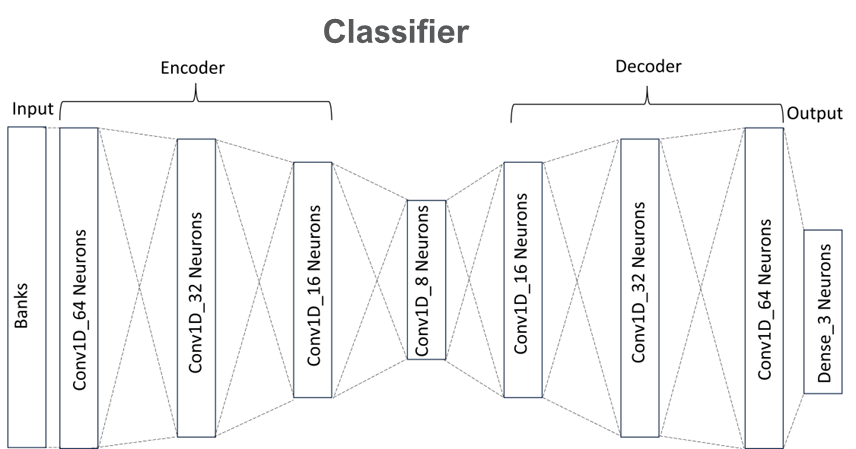}
\caption{Encoder-decoder structured 1D-CNN classifier}
\label{classifier}
\end{figure}

In our endeavor to advance fault diagnostics for control rod drive servomotors, our methodology seamlessly transitions into a comprehensive approach, integrating a structured encoder-decoder 1D-CNN classifier (Fig. \ref{classifier}). This section unfolds a strategic sequence of steps designed to proficiently diagnose faults within the operational parameters of control rod drives.

The process initiates with the selection of a first-order Keras optimizer, echoing the foundational importance observed in the fault isolation task. This optimizer, with default hyperparameters, significantly shapes the training of our encoder-decoder structured 1D-CNN classifier, acting as a core step for our fault diagnostics methodology.

The training phase assumes a central role, utilizing a substantial dataset comprising 70\% labeled faulty data. This extensive dataset exposes the classifier to a diverse array of fault scenarios, fostering its ability to discern patterns indicative of specific FMCRD faults. The encoder-decoder structure facilitates the extraction and representation of critical features essential for precise fault diagnosis.

Following the intensive training phase, the methodology progresses to critical validation and testing stages. The validation set, constituting 15\% of the samples, serves as a pivotal checkpoint for fine-tuning the model parameters and ensuring its robustness in recognizing fault patterns. Subsequently, the model undergoes a comprehensive test phase, classifying an unseen set of 15\% samples to evaluate its generalization capabilities and real-world applicability.

The entire fault diagnostics methodology is systematically repeated for each selected first-order Keras optimizer and operational property. This iterative approach facilitates a comprehensive comparative analysis, offering insights into the varied performances of classifiers and highlighting optimal configurations for distinct fault diagnosis scenarios.

\section{Results}
\label{sec:results}

In this section, the explained methods are implemented and different visualizations are employed for the different tasks.  The speed of convergence for the different optimizers was inferred from single runs of multiple epochs but predictive abilities were compared from multiple runs where each run has different initial parameters. The final validation loss for each run was compared for the isolation tasks whereas the minimum (best) validation loss for each run of multiple epochs was compared for diagnostics. The decision of the metric to compare for each task was made from assessing preliminary runs whose results were not reported in this work.

\subsection{Fault Isolation}
\label{subsec:result_isol}

This section employs bar charts to illustrate the average contribution of each control rod across multiple banks in order to identify and pinpoint faults. Specifically, if a control rod exhibits significantly higher average contributions than the other rods within the same bank, it implies that the fault originates from the particular rod and its associated FMCRD in that position. To elaborate further, in banks experiencing faults, certain rods are anticipated to display higher average contributions than their counterparts in corresponding positions in healthy banks. Notably, all faults in the studied banks are intentionally placed in position '10'. Consequently, the rods located in position 10 are expected to be the primary contributors to reconstruction errors in the faulty banks.

\begin{figure}[t!]
    \centering
    \includegraphics[width=0.44\textwidth, height = 5.5cm]{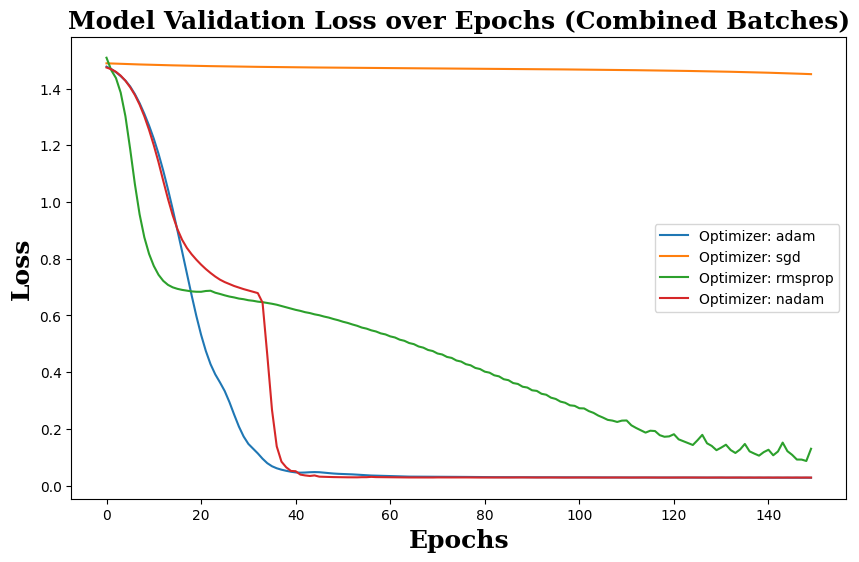}
    % \vspace{1em}
    \begin{minipage}[t]{0.45\textwidth}
        \raggedright
        \textbf{a)}\\
    \end{minipage}
    \hfill
    \vspace{1em}
    \includegraphics[width=0.44\textwidth]{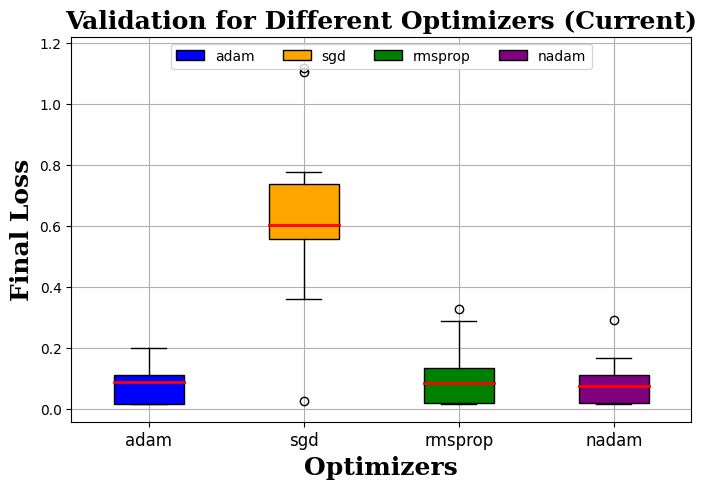}
    \begin{minipage}[t]{0.45\textwidth}
        \raggedright
        \textbf{b)}\\
    \end{minipage}
\caption{Validation loss profile from current monitoring. a) Single Run; b) Thirty runs.}
\label{current_val_loss_profile}
\end{figure}

Fig. \ref{current_val_loss_profile} shows the performances of the different optimizers when current is the monitored variable for the isolation task. When the single run of training with 150 epochs (Fig. \ref{current_val_loss_profile}a) is analyzed to measure the speed of convergence, the validation loss profiles show that Nadam was the quickest to reach its minimum possible loss with Adam as a close second. RMSProp in third has yet to converge completely but equally achieves low loss values. Over thirty runs (Fig. \ref{current_val_loss_profile}b) and ignoring outliers, Nadam has the lowest range of losses across runs with rmsprop and Adam coming close with equally low loss values. Considering how quickly Nadam converges before RMSProp and its low median final loss values across runs, Nadam is selected as the best performer in this case, with Adam as a close second.

\begin{figure}[t!]
    \centering
    \begin{minipage}[t]{\textwidth}
        \centering
        \includegraphics[width=0.45\textwidth]{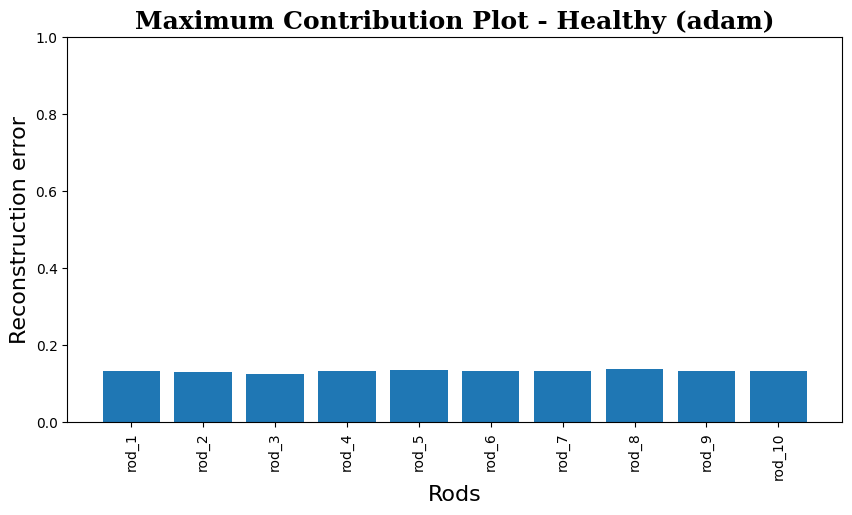}
        \vspace{0.5em}
        \raggedright
        \textbf{a)}
    \end{minipage}
    
    \begin{minipage}[t]{\textwidth}
        \centering
        \includegraphics[width=0.45\textwidth]{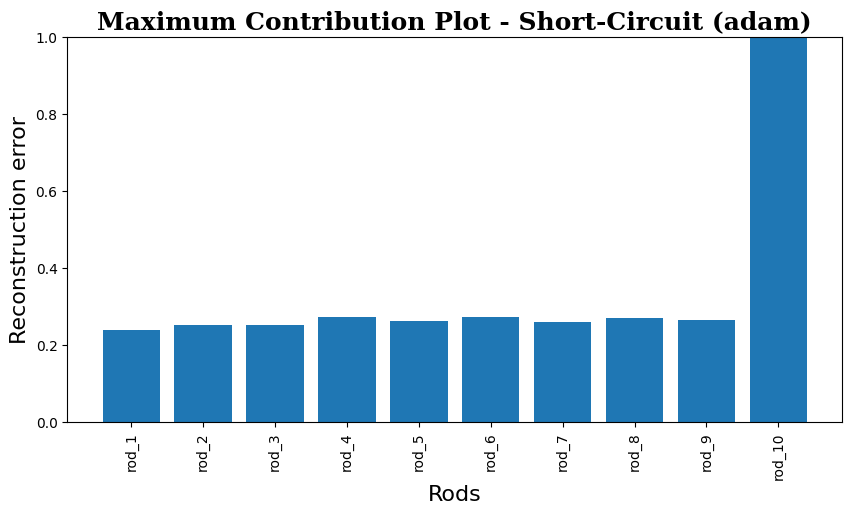}
        \vspace{1em}
        \raggedright
        \textbf{b)}
    \end{minipage}
    
    \begin{minipage}[t]{\textwidth}
        \centering
        \includegraphics[width=0.45\textwidth]{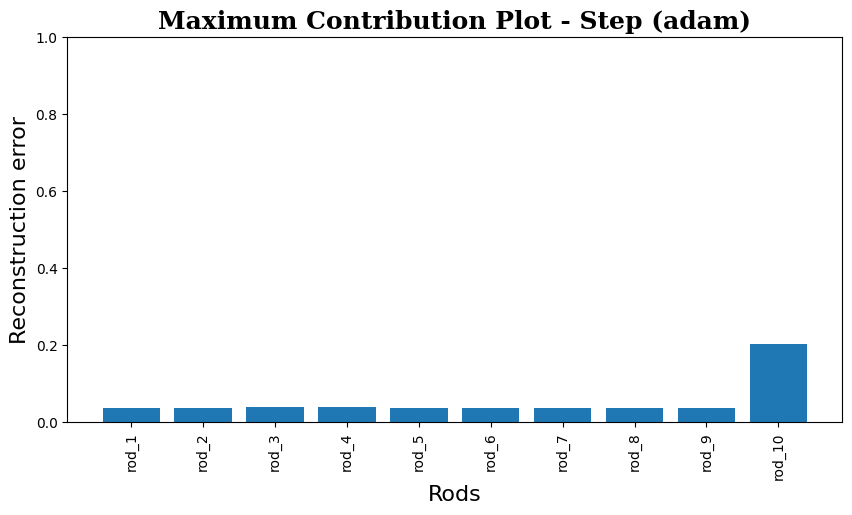}
        \vspace{0.5em}
        \raggedright
        \textbf{c)}
    \end{minipage}
    
    \begin{minipage}[t]{\textwidth}
        \centering
        \includegraphics[width=0.45\textwidth]{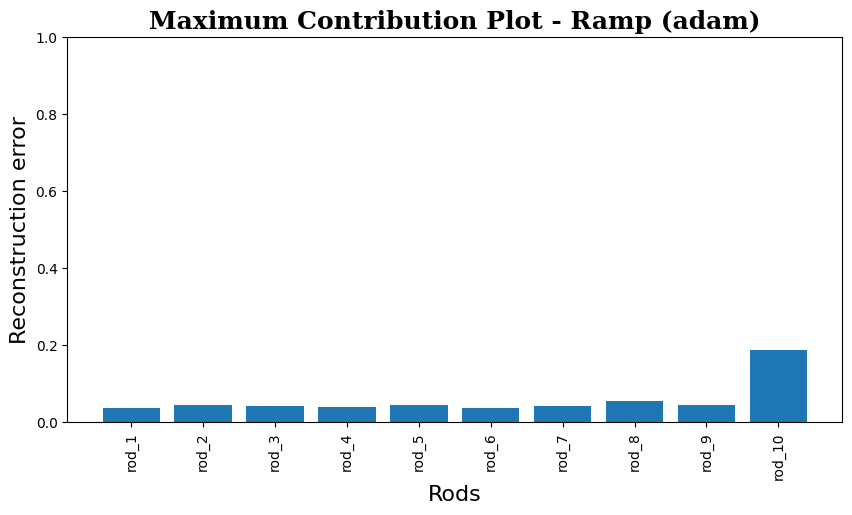}
        \vspace{0.5em}
        \raggedright
        \textbf{d)}
    \end{minipage}
    
    \caption{Fault isolation with current monitoring. a) - healthy; b) - short circuit fault; c) - step fault; d) - ramp fault.}
    \label{isol_current_results}
\end{figure}

During testing, all the general models (RMSProp, Nadam, and Adam) were able to detect and correctly isolate the faults in the 10$^{th}$ rod of the three different faulty banks. However, the mechanical faults were harder to isolate as displayed in Fig. \ref{isol_current_results}  for the RMSProp optimizer.

\begin{figure}[t!]
    \centering
    \includegraphics[width=0.44\textwidth, height = 5.5cm]{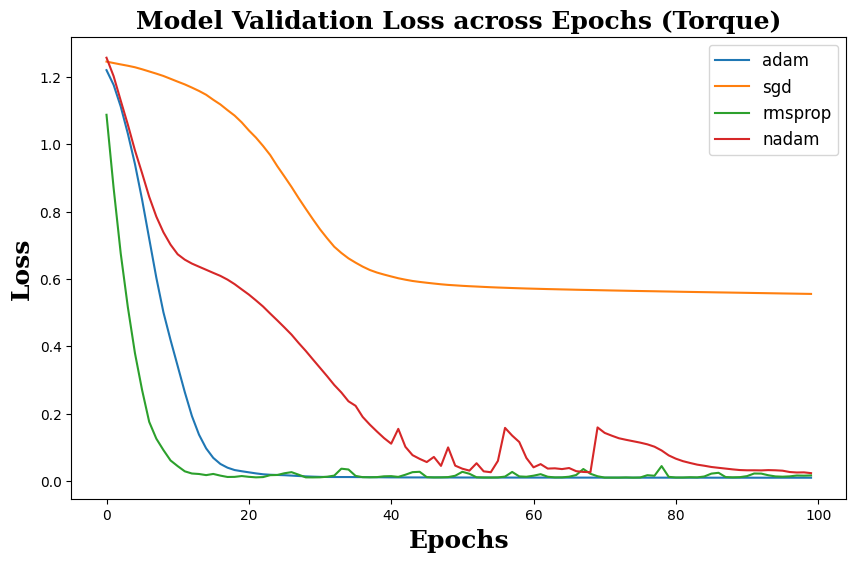}
    % \vspace{1em}
    \begin{minipage}[t]{0.45\textwidth}
        \raggedright
        \textbf{a)}\\
    \end{minipage}
    \hfill
    \vspace{1em}
    \includegraphics[width=0.44\textwidth]{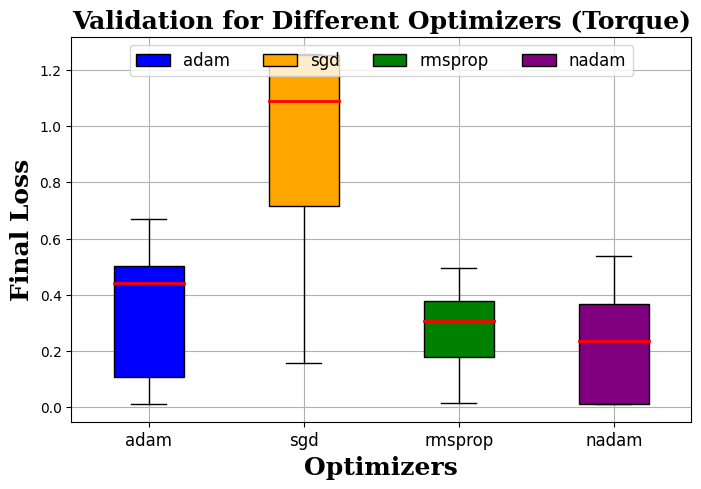}
    \begin{minipage}[t]{0.45\textwidth}
        \raggedright
        \textbf{b)}\\
    \end{minipage}
\caption{Validation loss profile from torque monitoring. a) Single Run; b) Thirty runs.}
\label{torque_val_loss_profile}
\end{figure}

\begin{figure}[t!]
    \centering
    \begin{minipage}[t]{\textwidth}
        \centering
        \includegraphics[width=0.45\textwidth]{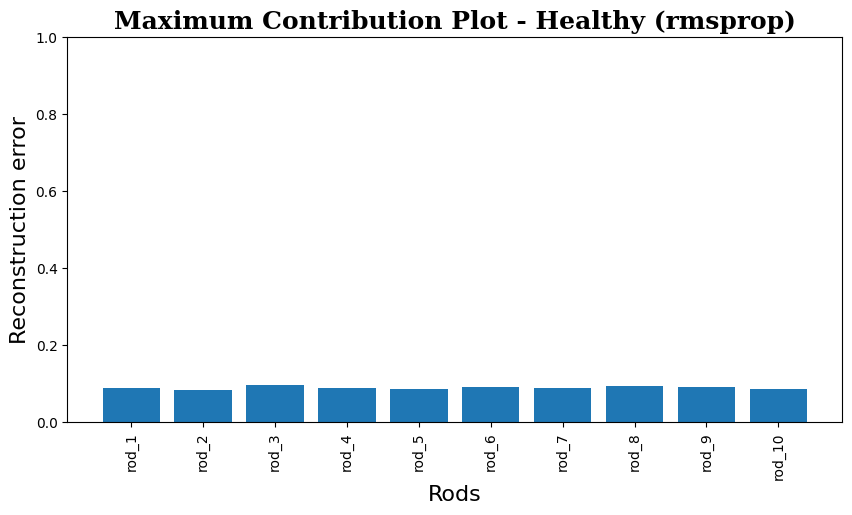}
        \vspace{0.5em}
        \raggedright
        \textbf{a)}
    \end{minipage}
    \hfill
    \begin{minipage}[t]{\textwidth}
        \centering
        \includegraphics[width=0.45\textwidth]{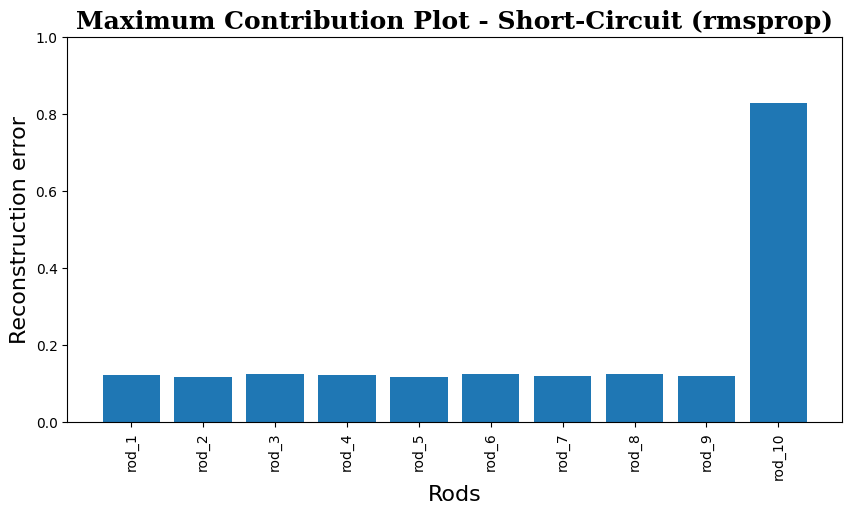}
        \vspace{1em}
        \raggedright
        \textbf{b)}
    \end{minipage}
    \begin{minipage}[t]{\textwidth}
        \centering
        \includegraphics[width=0.45\textwidth]{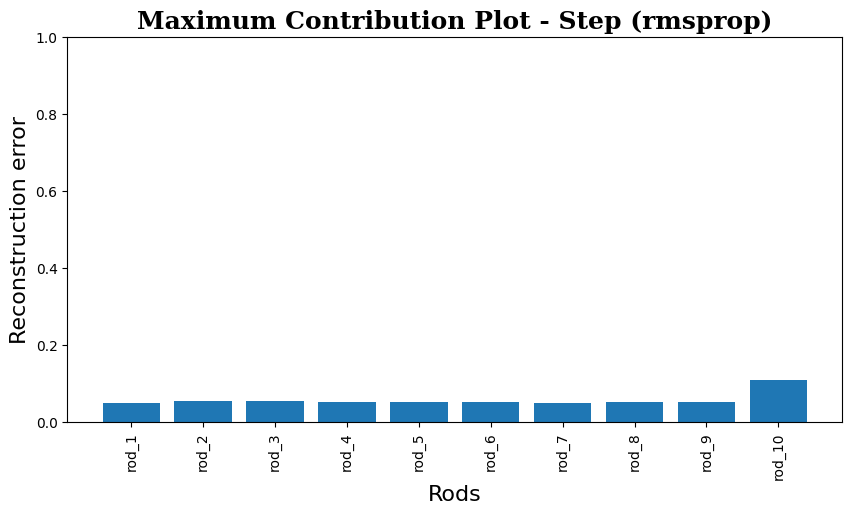}
        \vspace{0.5em}
        \raggedright
        \textbf{c)}
    \end{minipage}
    \hfill
    \begin{minipage}[t]{\textwidth}
        \centering
        \includegraphics[width=0.45\textwidth]{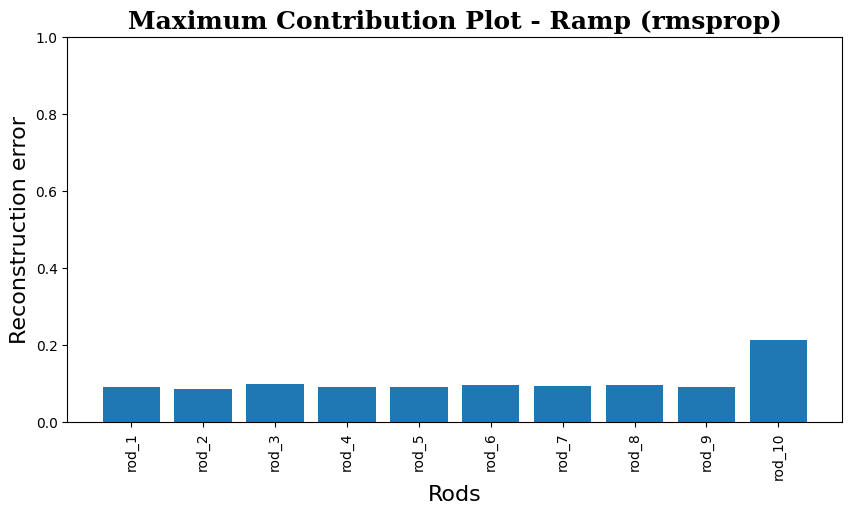}
        \vspace{0.5em}
        \raggedright
        \textbf{d)}
    \end{minipage}
    
    \caption{Fault isolation with torque monitoring. a) - healthy; b) - short circuit fault; c) - step fault; d) - ramp fault.}
    \label{isol_torque_results}
\end{figure}

In the case of isolation from torque monitoring, Fig. \ref{torque_val_loss_profile}a, the single run of 100 epochs, shows rmsprop as the quickest to converge whereas the thirty runs plot (Fig. \ref{torque_val_loss_profile}b) also suggests that rmsprop guarantees low loss values but Nadam gave the best median and loss range. Here, Nadam is selected as the overall best performer with test results similar to the previous case (see Fig. \ref{isol_torque_results}).

Overall, the adaptive and more general optimizers can adequately detect and isolate the FMCRD faults for the three investigated fault types with either monitored property but rmsprop and Nadam outperformed Adam, especially with torque monitoring.

\subsection{Fault Diagnostics}
\label{subsec:result_diag}

Since fault diagnostics is a classification task, confusion matrices are used to show the test results in this section. Optimization algorithms are compared with loss values and convergence speed as in the preceding section.

\begin{figure}[t!]
    \centering
    \includegraphics[width=0.44\textwidth, height = 5.5cm]{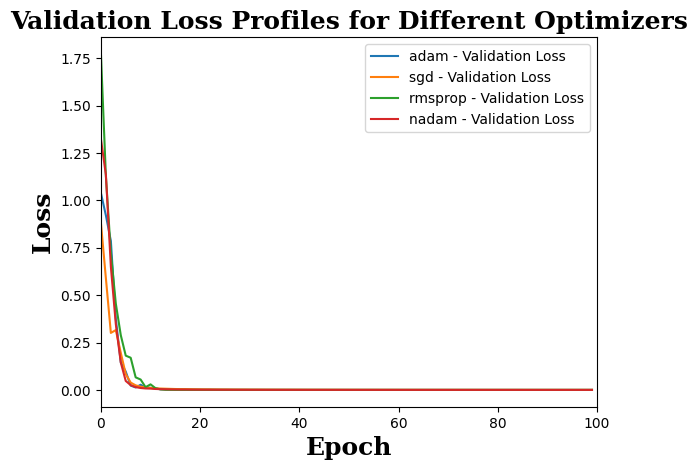}
    % \vspace{1em}
    \begin{minipage}[t]{0.45\textwidth}
        \raggedright
        \textbf{a)}\\
    \end{minipage}
    \hfill
    % \vspace{0.5em}
    \includegraphics[width=0.44\textwidth]{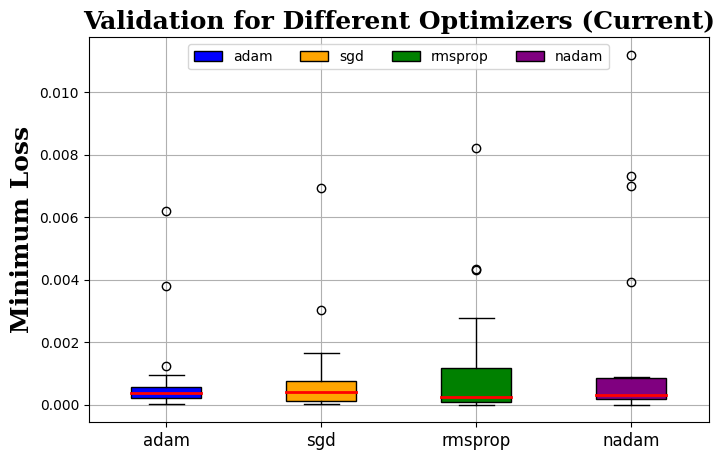}
    \begin{minipage}[t]{0.45\textwidth}
        \raggedright
        \textbf{b)}\\
    \end{minipage}
\caption{Validation loss profile from current monitoring. a) Single Run; b) Twenty runs.}
\label{current_val_loss_profile_diag}
\end{figure}

\begin{figure}[h!]
\centering
\includegraphics[width=0.45\textwidth]{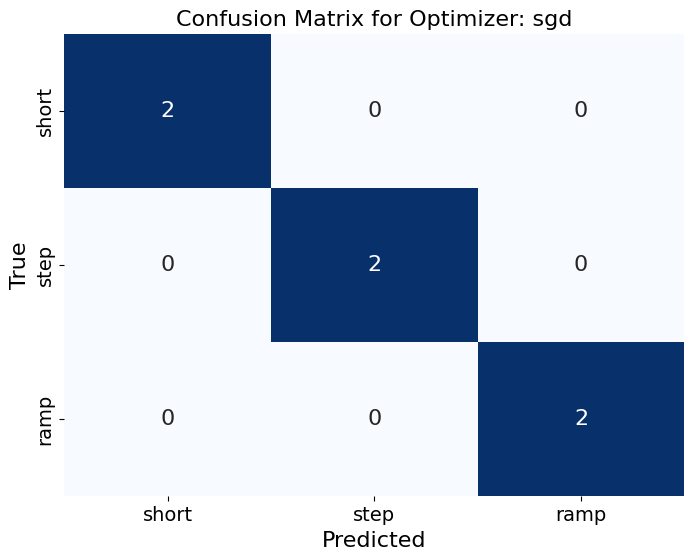}
\caption{Diagnostics test results with current monitoring}
\label{curr_diag_test}
\end{figure}

Fig. \ref{current_val_loss_profile_diag} shows the diagnostics results for current monitoring and as seen, the speed of convergence is similar for all optimizers and the prediction performances across twenty runs also appear close. However, Adam is selected because of its higher confidence, ignoring the outlier values. The confusion matrix for the test data (Fig. \ref{curr_diag_test}) shows 100\% identification of all faults with SGD but all optimizers produced the same test result.

\begin{figure}[t!]
    \centering
    \includegraphics[width=0.44\textwidth, height = 5.5cm]{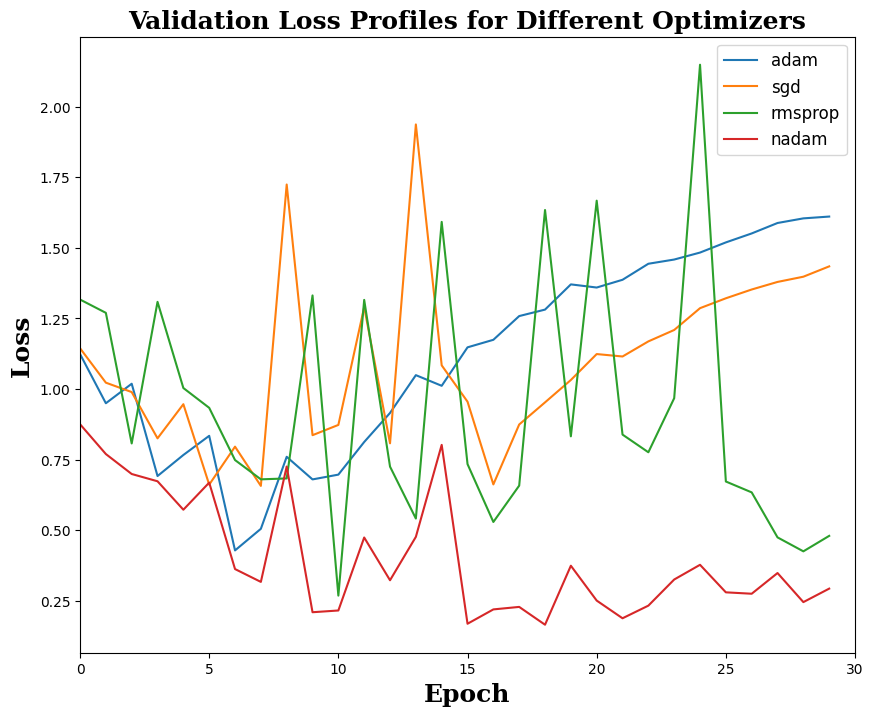}
    % \vspace{1em}
    \begin{minipage}[t]{0.45\textwidth}
        \raggedright
        \textbf{a)}\\
    \end{minipage}
    \hfill
    \vspace{1em}
    \includegraphics[width=0.44\textwidth]{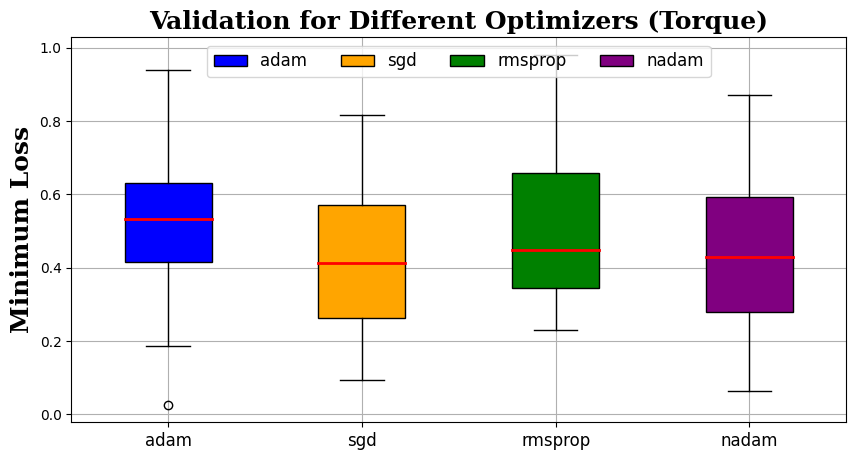}
    \begin{minipage}[t]{0.45\textwidth}
        \raggedright
        \textbf{b)}\\
    \end{minipage}
\caption{Validation loss profile from torque monitoring. a) Single Run; b) Twenty runs.}
\label{torque_val_loss_profile_diag}
\end{figure}

\begin{figure}[h!]
\centering
\includegraphics[width=0.45\textwidth]{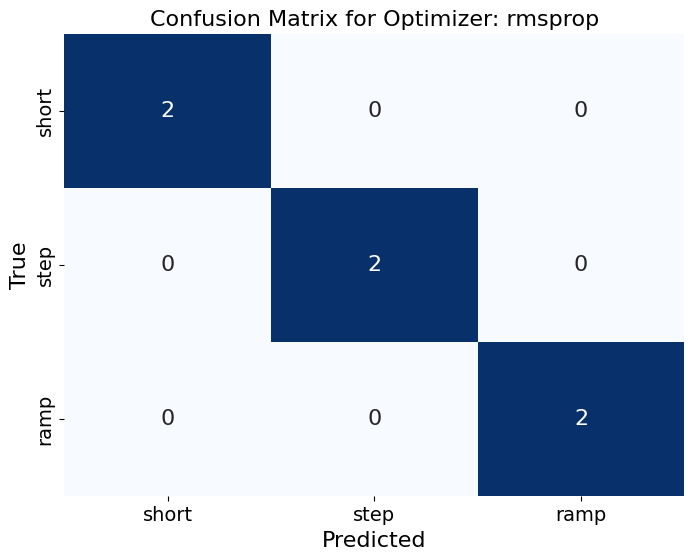}
\caption{Diagnostics test results with torque monitoring}
\label{tor_diag_test}
\end{figure}

For torque (Fig. \ref{torque_val_loss_profile_diag}), the single run shows signs of overfitting with all optimizers but the lowest validation loss was achieved with Nadam around the 15$^{th}$ epoch. Over multiple runs, however, a similar low median of minimum loss across epochs was achieved with SGD, rmsprop, and Nadam but SGD had the lowest median leading to its selection as the best. Testing with the best models, all optimizers achieved 100\% test accuracies (see Fig. \ref{tor_diag_test} for rmsprop) but from the multiple runs, it can be deduced that Adam would find it most challenging to replicate that result.

Overall, both torque and current were useful properties for the investigated tasks. This aligns with what is seen in the literature since the most used diagnostics method for nuclear reactors (MCSA) relies on monitoring and analyzing electric current signals \cite{mcsa3}. In this research, however, predefined behaviors of the current signal under the investigated faults were not required. Moreover, in an electric motor, the relationship between electric current and electromagnetic torque is described by the motor's torque equation (equation \ref{eq:torqu_equation}) \cite{torque_current}, offering a possible explanation for the effectiveness of torque.

\begin{equation}
\label{eq:torqu_equation}
T_{\text{em}} = K_{\text{torque}} \cdot \Phi \cdot I_a 
\end{equation}

Here, \(T_{\text{em}}\) represents the electromagnetic torque, \(K_{\text{torque}}\) is the torque constant, which characterizes the motor's torque-producing ability. \(\Phi\) denotes the magnetic flux, a quantity dependent on the motor's design and the field current. \(I_a\) is the armature current, signifying the electric current supplied to the motor's armature winding. As the armature current (\(I_a\)) increases, the electromagnetic torque (\(T_{\text{em}}\)) produced by the motor also increases approximately linearly, provided that the magnetic flux (\(\Phi\)) remains constant.

\begin{table}
    \centering
    \caption{Summary of Performances}
    \label{tab:performance_summary}
    \begin{tabular}{p{1.7cm} p{1.75cm} p{1.75cm} p{1.93cm}} % Adjust column widths as needed
        \toprule
        Task & Monitored Property & Best Optimizer & Worst Optimizer \\
        \midrule
        Isolation & Current & Nadam & SGD \\
        Isolation & Torque & Nadam & SGD \\
        Diagnostics & Current & Adam & None \\
        Diagnostics & Torque & SGD & Adam \\
        \bottomrule
    \end{tabular}
\end{table}

Table \ref{tab:performance_summary} shows the summary of the results so far and Nadam was mostly preferred for the different highlighted reasons. It should be noted that in this work, the convergence rate was a secondary metric to validation loss distribution because it depended on a single run. Hence, the convergence rate was only considered when there was no clear difference in validation loss performance (Fig. \ref{current_val_loss_profile}b). In addition, SGD performed poorly for the isolation tasks but showed comparable performances for diagnostics.

Across both isolation and diagnostics tasks, Nadam produced the best and most generalizable predictors. This is important because Adam is usually the default optimizer for most practitioners \cite{moolayil2019learn}. A possible reason for the overall performance of Nadam is that its adaptive gradient component which allows it to adjust step size for each parameter and enables it to better handle various data distributions and complexities. Additionally, its Nesterov accelerated gradient component, allows it to anticipate future gradient updates, leading to smoother convergence and helping escape local minima more efficiently. 

SGD performed better for classification tasks than reconstruction tasks with autoencoders possibly due to the nature of the optimization process and the challenges associated with each task. In classification tasks, the goal is to learn a decision boundary between different classes, and SGD's simplicity and efficiency in updating model parameters based on the gradients of the loss function make it effective for this purpose. On the other hand, in reconstruction tasks with autoencoders, the objective is to accurately reconstruct input data, which involves capturing intricate relationships and dependencies among variables. The fixed learning rate of SGD, without adaptive adjustments, can hinder its ability to navigate the complex and high-dimensional loss landscape associated with autoencoder-based reconstruction, potentially leading to suboptimal convergence and performance \cite{sgd}. Adaptive optimizers are often more suitable for autoencoder-based tasks as they dynamically adjust learning rates based on the historical behavior of each parameter, facilitating smoother convergence in intricate reconstruction scenarios \cite{autoencoderRecon}.

\subsection{Effect of Runs}
\label{subsec:result_multiple_runs}

\begin{figure}[t]
    \centering
    \includegraphics[width=0.44\textwidth, height = 5.5cm]{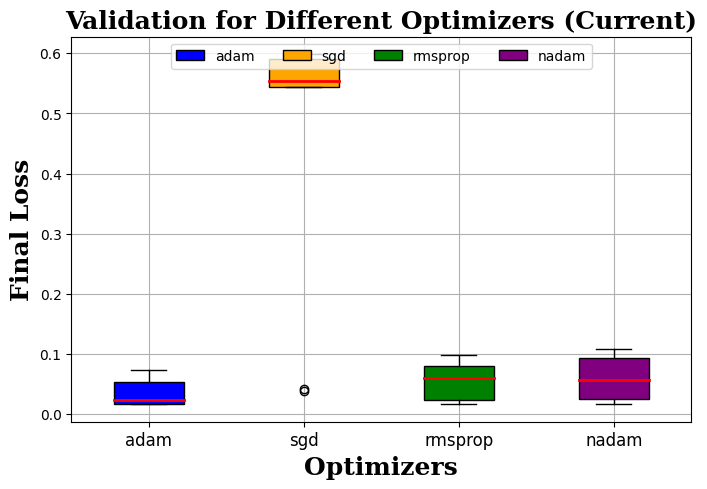}
    % \vspace{1em}
    \begin{minipage}[t]{0.45\textwidth}
        \raggedright
        \textbf{a)}\\
    \end{minipage}
    \hfill
    \vspace{0.5em}
    \includegraphics[width=0.44\textwidth, height = 5.5cm]{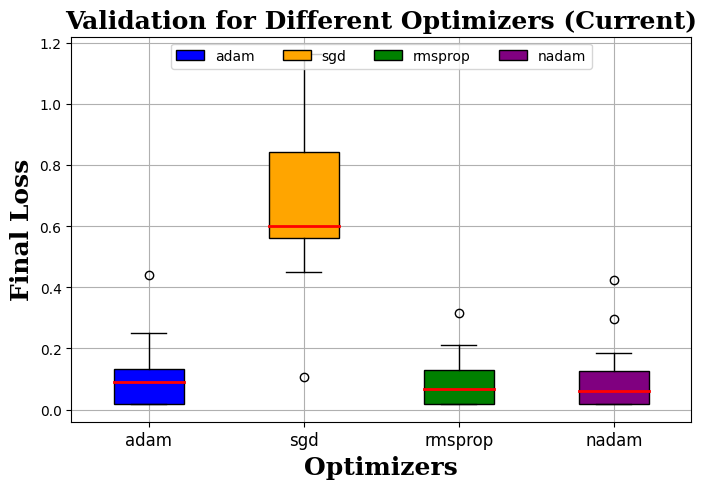}
    \begin{minipage}[t]{0.45\textwidth}
        \raggedright
        \textbf{b)}\\
    \end{minipage}
\caption{Validation loss profile from current monitoring for isolation with the different number of runs. a) Ten Runs; b) Forty runs.}
\label{muti_runs_current_isol}
\end{figure}

A single run might not capture the variability inherent in complex tasks, potentially leading to misleading conclusions. Conversely, an excessive number of runs may incur unnecessary computational costs. Thus, finding the optimal balance in the number of runs is imperative to strike a balance between computational efficiency and result reliability, allowing for more informed conclusions in empirical model comparisons. Consequently, the effect of the number of runs for performance evaluation was investigated with two workloads: isolation with current; and diagnostics with torque. For the first workload, with the single run in Fig. \ref{current_val_loss_profile}a, Nadam and Adam achieved similar loss values, but with ten runs as shown in Fig. \ref{muti_runs_current_isol}a, Adam became the favorite in terms of median final validation loss. With forty runs (Fig. \ref{muti_runs_current_isol}b), Nadam provided a better median loss value with Adam and rmsprop not too far behind. 

\begin{figure}[h!]
    \centering
    \begin{minipage}[t]{\textwidth}
        \centering
        \includegraphics[width=0.45\textwidth]{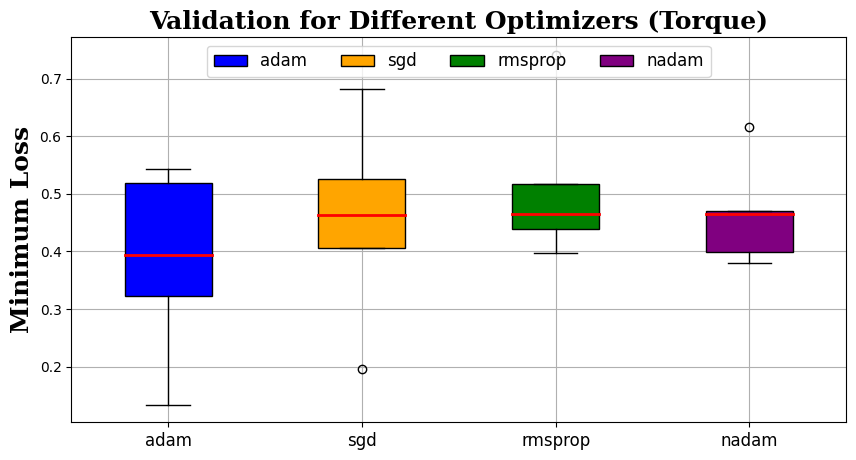}
        \vspace{1em}
        \raggedright
        \textbf{a)}
    \end{minipage}
    \begin{minipage}[t]{\textwidth}
        \centering
        \includegraphics[width=0.45\textwidth]{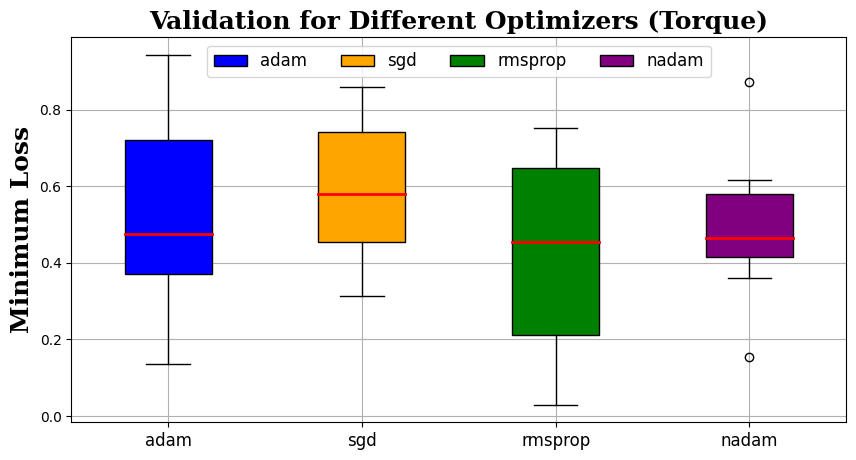}
        \vspace{0.5em}
        \raggedright
        \textbf{b)}
    \end{minipage}
    \hfill
    \begin{minipage}[t]{\textwidth}
        \centering
        \includegraphics[width=0.45\textwidth]{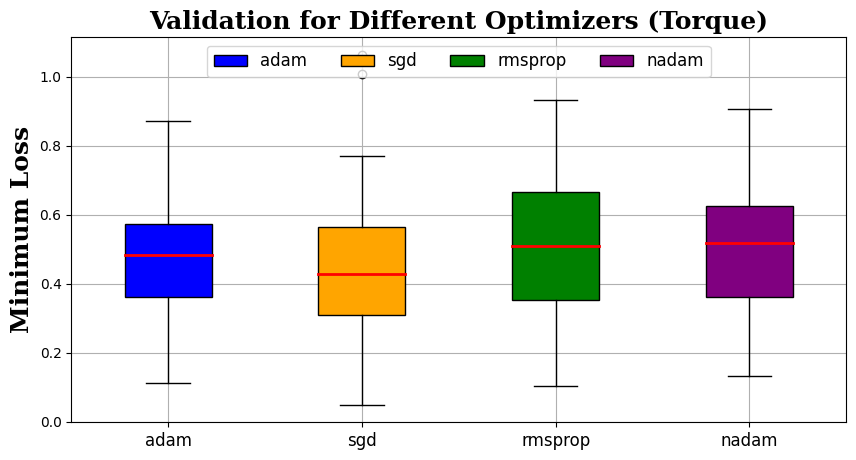}
        \vspace{0.5em}
        \raggedright
        \textbf{c)}
    \end{minipage}
    
    \caption{Best validation loss of the diagnostics classifier from torque monitoring with different numbers of runs. a) - Five runs; b) - Ten runs; c) - Forty runs.}
    \label{muti_runs_torque_diag}
\end{figure}
For the second workload, with one run, Nadam achieved the lowest minimum loss (Fig. \ref{torque_val_loss_profile_diag}a). With five runs, the best median was seen with Adam in Fig. \ref{muti_runs_torque_diag}a. Ten runs favored rmsprop (Fig. \ref{muti_runs_torque_diag}b), whereas forty runs suggested SGD as the best (Fig. \ref{muti_runs_torque_diag}c). 

From the analyses in this section, two deductions can be made. First, the best optimizer choice for any workload depends on the number of runs. Second, the optimal number of runs in this work was thirty for the isolation task and twenty for the diagnostics task.    
Comparing Fig. \ref{current_val_loss_profile}b and Fig. \ref{muti_runs_current_isol}b respectively representing thirty runs and forty runs of the same workload as well as Fig. \ref{torque_val_loss_profile_diag}b and Fig. \ref{muti_runs_torque_diag}c respectively representing twenty runs and forty runs of similar workloads, it is seen that the median values of the different optimizers did not significantly change and the best and worst optimizers remained the same. This justifies the number of runs used for comparisons in the preceding sections (\ref{subsec:result_isol} and \ref{subsec:result_diag}).

While the proposed classification algorithm exhibits promising results within the simulated environment, its feasibility in real-world experimental data warrants thorough consideration. Although some measurement noise was added to the simulation to make it more realistic, transitioning from simulated to real-world data introduces complexities such as uncertainties, and unforeseen operating conditions. Additionally, the precise replication of fault scenarios encountered in real-world scenarios within simulations may be challenging. Hence, the algorithm's robustness and adaptability to real-world dynamics need validation through comprehensive experimental testing. Real-world experimental data collection from the BWRX-300 SMR, incorporating diverse operating conditions and fault scenarios when they become operational, would provide invaluable insights into the algorithm's performance. Moreover, the algorithm's sensitivity to sensor placement and the availability of sensor data in practical SMR setups must be assessed. Addressing these challenges and validating the algorithm's efficacy in real-world scenarios is crucial for ensuring its successful deployment in enhancing fault detection and diagnostics in SMR control systems.

% \newpage
\section{Conclusion}
\label{sec:conclusion}
Firstly, it was shown from the test results that fault isolation and diagnostics of the BWRX-300 FMCRD can be consistently achieved with the proposed methods. Secondly, from the brief comparison of the monitored properties for each task, the electric current signal appears to be a silver bullet useful for all the investigated tasks. Therefore, a sensor that measures current or proxy properties should be strategically placed in relevant areas in the BWRX-300 SMR to swiftly and accurately detect, isolate, and diagnose its FMCRD faults to potentially ultimately drive down its O\&M costs. Thirdly, based on the selection criteria, dataset, and methods applied, Nadam was the best optimizer for most of the investigated tasks. Lastly, it was also shown that empirical comparison of optimizers should treat the number of runs for model assessment as hyperparameters to be optimized.  

As an extension to this work, tuning the hyperparameters of each optimization algorithm before comparing the best models should be considered. Furthermore, setting a target validation error and measuring the speed of convergence to that target point across multiple runs for different optimizers could be examined.

\section*{Acknowledgment}

This material is based upon work supported by the Department of Energy under Award Number DE-NE0009278.

This report was prepared as an account of work sponsored by an agency of the United States Government. Neither the United States Government nor any agency thereof, nor any of their employees, makes any warranty, express or implied, or assumes any legal liability or responsibility for the accuracy, completeness, or usefulness of any information, apparatus, product, or process disclosed, or represents that its use would not infringe privately owned rights. Reference herein to any specific commercial product, process, or service by trade name, trademark, manufacturer, or otherwise does not necessarily constitute or imply its endorsement, recommendation, or favoring by the United States Government or any agency thereof. The views and opinions of authors expressed herein do not necessarily state or reflect those of the United States Government or any agency thereof.



%\bibliographystyle{IEEEtran}
% \newpage
\newpage
\bibliographystyle{IEEEtran}%alpha
% \newpage
\bibliography{References}



% \end{thebibliography}

\end{document}